# Thermodynamics of boson and fermion systems with fractal distribution functions

Marcelo R. Ubriaco*
*Laboratory of Theoretical Physics*
*Department of Physics*
*University of Puerto Rico*
*P. O. Box 23343, Río Piedras*
*PR 00931-3343, USA*

Starting with the fractal inspired distribution functions for Maxwell-Boltzmann, Bose-Einstein and Fermi systems, as reported by F. Büyükkiliç and D. Demirhan, we obtain the corresponding probability distributions and study their thermodynamic behavior. We compare our results with those corresponding to ideal gases ($q = 1$), and Bose-Einstein and Fermi systems with quantum group symmetry. In particular, we show that the hamiltonian that gives the Bose-Einstein generalized distribution function can be interpreted as a $q$-deformation of the ideal gas hamiltonian.

## I. INTRODUCTION

As is well known from the theory of fractals, given the statistical weight $\Omega(q, r)$ of a system and the resolution $r$, the fractal dimension is defined as the exponent $d = D_q$ which will render the product $\lim_{r\to 0}\Omega(q, r)r^d$ finite. Since the entropy $S(q, r)$ is proportional to $\ln\Omega(q, r)$, the relation between the entropy $S(q, r)$ and the generalized dimensions $D_q$ is given by

$$D_q = -\lim_{r\to 0}\frac{S(q, r)}{\ln r}, \tag{1}$$

where the $q$ parameter stands for the moment order. The entropy function will depend on a set of probabilities $p_i$ that a random variable will be found into the ith bin of size $r$. Based on these definitions and by use of the Boltzmann's $H$-theorem, the generalized entropy and distribution functions for classical and quantum gases were calculated in Ref [1]. The average number of particles with energy $\varepsilon$ were shown to be given by

$$\langle n(\varepsilon)\rangle = \frac{1}{[1 + \beta(q - 1)(\varepsilon - \mu)]^{1/(q-1)} + a}, \tag{2}$$

where $a = 0$ for the classical case and the values $a = -1, +1$ correspond to Bose-Einstein and Fermi-Dirac cases respectively. For $q = 1$, Eq.(2) becomes the standard textbook result for classical and quantum gases. Therefore, for arbitrary $q$ and within the context of statistical mechanics, the corresponding formalism can be understood as a generalization of Boltzmann-Gibbs statistics. Two examples of generalizations of the concept of entropy based on ideas from the theory of fractals are the functions introduced by Rényi [2] and Tsallis [3], which are given respectively as

*E-mail:ubriaco@ltp.upr.clu.edu

$$S_q^R = \frac{k}{q-1} \ln \sum_R p_R^q, \tag{3}$$

$$S_q = \frac{k}{q-1} \left( 1 - \sum_R p_R^q \right), \tag{4}$$

where $p_R$ is the probability for the ensemble to be in the state $R$. It is clear that both entropy functions become the Shannon entropy function $S = -k \sum_R p_R \ln p_R$ as $q \to 1$. A discussion on the irreversible character of these entropies can be found in Refs. [4]. In particular, the entropy function $S_q$ has all the properties of the Shannon entropy except that of additivity. In fact, given two independent systems $\Sigma$ and $\Sigma'$ the function $S_q$ satisfies

$$\frac{S_q^{\Sigma \cup \Sigma'}}{k} = \frac{S_q^{\Sigma}}{k} + \frac{S_q^{\Sigma'}}{k} + (1-q) \frac{S_q^{\Sigma}}{k} \frac{S_q^{\Sigma'}}{k}, \tag{5}$$

which becomes additive for $q = 1$. Therefore, Tsallis' formulation provides a framework to deal with the nonextensive properties of certain physical systems. Numerous applications of this formalism [5], and similar formulations appeared in the literature [6]. The equilibrium probability distribution $p_R$ in the grand canonical ensemble is written as

$$p_R = \frac{\left[1 + \beta(q-1)(E_R - \mu N)\right]^{1/(1-q)}}{Z_q}, \tag{6}$$

with the partition function

$$Z_q = \sum_R \left[1 + \beta(q-1)(E_R - \mu N)\right]^{1/(1-q)}, \tag{7}$$

and the total energy $E_R = \sum_j n_j \varepsilon_j$. Thermodynamic stability is achieved by defining a $q$-averaging [7]

$$\langle E \rangle = \sum_R p_R^q E_R, \tag{8}$$

leading to expressions of the type

$$\langle E \rangle = -\frac{\partial}{\partial \beta} \frac{Z_q^{1-q} - 1}{1-q}. \tag{9}$$

The fact that the partition function in Eq. (7) does not factorize in independent modes is a consequence of the nonextensive characteristics, like those involving long range interactions, of a physical system. Due to the mathematical complexity of dealing with equations like (7), calculations in this direction have been performed [8] for values of $q \approx 1$.

The distribution functions in Eq. (2) and nonextensive thermodynamics are not unrelated. In fact, these fractal distributions have also been obtained [9] by considering the case of a dilute gas and approximating the partition function in Eq. (7) with a factorized partition function. This approximation to the nonextensive thermodynamics of Ref. [3], which ignore the correlations between the particles, has been shown [10] to be a good one outside a certain temperature interval. This time interval shifts to higher values of $T$ when either the number of particles or the number of energy levels are increased. Some applications of the dilute gas approximation as an approximate approach to nonextensive thermodynamics can be found in Refs. [11].

It has been also pointed out [12] that nonextensive thermodynamics could be understood in terms of $q$-deformations. The fact behind this possible connection between two apparently unrelated subjects resides basically in the observation that a $q$-number $[x] \equiv (q^x - 1)/(q - 1)$ has the pseudoadditivity property of the type displayed in Eq. (5), and thus the entropy $S_q$ can be defined in terms of the $q$-derivative $D_x = x^{-1}(1 - q^{xd/dx})/(1 - q)$ acting on the probability distribution. The theory of $q$-analysis was formulated at the beginning of this century and it serves as an analytical framework to study basic hypergeometric series [13]. The study of the so-called $q$-series dates back to the times of Euler, and it has played an important role in the theory of partitions [14] and more recently in quantum group theory [15]. Although nonextensive thermodynamics does not embody any quantum group structure, it could be somehow related to the theory of quantum groups. The link to quantum groups is through $q$-analysis. As is well known, noncommutative calculus arises as a set of algebraic relations which are covariant under quantum group transformations [16]. It has been shown [17] that differential operators in noncommutative calculus has a representation in classical space in terms of $q$-derivatives times scaling operators. Therefore, although the link between nonextensive thermodynamics and quantum groups seems to be rather marginal, these two distinct subjects share a common language which is that of $q$-analysis. In addition, it has been shown [18] that generators of multifractal sets which are self-similar under $q$-scalings can be identified with the $q$-derivatives $D_x$.

Then, according to the work of several authors, the fractal distribution functions in Eq. (2) are those corresponding to a dilute gas approximation of nonextensive thermodynamics, and the formalism of Tsallis thermodynamics seems to be related to the theory of quantum groups through the use of $q$-analysis. Therefore, it is natural to investigate whether the thermodynamic functions arising from the fractal distribution functions in Eq.(2) are somehow similar to the thermodynamic behavior of systems with quantum group symmetry. The meaning of the parameter $q$ should be understood within the appropriate context. In the theory of fractals $q$ stands for the moment order that defines the fractal measure [19], in the theory of quantum groups $q$ parametrizes a deformation from classical Lie group symmetry, and in Tsallis thermodynamics $q$ parametrizes the departure from extensive thermodynamics. In this work, the most appropriate interpretation of $q$ is as a parameter that measures the departure of the corresponding statistical mechanics from the Boltzmann-Gibbs formulation.

Our motivation in studying this system is twofold. First, as discussed above, it relates to the interesting fact that the fractal distributions functions approximate well, within certain limits, nonextensive behavior. Second, it is of theoretical interest to study a thermodynamic system whose probability density generalizes Boltzmann-Gibbs statistics. Our calculations will also indicate whether the corresponding behavior from these generalized statistics shares some properties with the thermodynamics of quantum gases with quantum group symmetry.

As our starting point we consider the two following basic requirements. In order that the applicability of the thermodynamic limit makes sense we consider a factorized probability density. The formulation should also be consistent with the classical limit.

This paper is organized as follows. In Section 2 we obtain the probability density and density operators for Maxwell-Boltzmann and quantum statistics respectively that lead to the particle distributions functions in Eq.(2). For the case of Maxwell-Boltzmann statistics, we find the correct way of calculating the thermodynamic functions such that the expected classical behavior occurs. For quantum statistics, with use of the standard operator formalism we derive the boson and fermion Hamiltonians that leads to the occupation number functions in Eq.(2), and based on this we study the thermodynamics of these systems in the thermodynamic limit. In particular, for the Bose-Einstein case we calculate the dependence of the internal energy, heat capacity and entropy on the parameter $q$ and the temperature. For quantum statistics we compare our results with the thermodynamics of quantum group invariant systems as reported in Refs. [20,21]. In Section 3 we summarize our results.

## II. THERMODYNAMICS OF BOSON AND FERMION SYSTEMS

### A. Maxwell-Boltzmann case

The particle distribution for the Maxwell-Boltzmann case is given by Eq. (2) for $a = 0$.

$$\langle n_l \rangle = \rho_l^{-1}, \tag{10}$$

where

$$\rho_l = \left(1 + (q-1)\beta(\varepsilon_l - \mu)\right)^{1/(q-1)}. \tag{11}$$

We define the probability density $\rho$ according to the equation

$$\rho = \frac{1}{Z_{MB}} \prod_{l=0} \frac{1}{n_l!} \rho_l^{-n_l}, \tag{12}$$

with the corresponding partition function

$$\begin{aligned} Z_{MB} &= \prod_{l=0} \sum_{n_l=0} \frac{1}{n_l!} \rho_l^{-n_l}, \\ &= \prod_{l=0} e^{\rho_l^{-1}}. \end{aligned} \tag{13}$$

A simple check shows that from the definition of the average total number of particles $\langle N \rangle$

$$\begin{aligned} \langle N \rangle &= \sum_{l=0} \frac{\sum_{n_l=0} (1/n_l!) n_l \rho^{-n_l}}{\sum_{n_l=0} (1/n_l!) \rho^{-n_l}}, \\ &= -\sum_{l=0} \rho^{q-1} \frac{1}{\beta} \frac{\partial \ln Z_{MB}}{\partial \varepsilon_l}, \end{aligned} \tag{14}$$

the probability density in Eq.(12) leads to the particle distribution in Eq.(10). In the thermodynamic limit we rewrite Eq. (14) in terms of the integral

$$\langle N \rangle = \frac{4\pi V}{h^3} \left( \frac{2m}{\beta(q-1)} \right)^{3/2} \int_0^\infty \frac{x^2 dx}{\left(1 + x^2 - (q-1)\beta\mu\right)^{1/(q-1)}}. \tag{15}$$

For further purposes we wish to consider the integral

$$I_\delta = \int_0^\infty \frac{x^\delta dx}{(1+x^2-(q-1)\beta\mu)^{1/(q-1)}}. \tag{16}$$

The requirement of finiteness for $I_\delta$ restricts the values of $q$ to the interval $q \in [1,(4+\delta)/(2+\delta)]$. Defining a new variable $w = 1+x^2-(q-1)\beta\mu$ and expanding the integrand in powers of $((q-1)\beta\mu - 1)/w$ we find that $I_\delta$ is given by the expression

$$I_\delta = -\frac{1}{2}\frac{1}{(1-(q-1)\beta\mu)^{1/(q-1)-(\delta+1)/2}} S_\delta, \tag{17}$$

where $S_\delta$ is a series independent of $\beta\mu$ given by

$$S_\delta = \frac{1}{-1/(q-1)+(\delta+1)/2} + \sum_{m=1}^{\infty} \frac{(-1)^m}{m!}\left(\frac{\delta-1}{2}\right)...\left(\frac{\delta+1}{2}-m\right)\frac{1}{-1/(q-1)-m+(\delta+1)/2}. \tag{18}$$

Therefore, for Eq.(15) we write

$$\langle N\rangle = \frac{-2\pi V}{h^3}\left(\frac{2m}{\beta(q-1)}\right)^{3/2}\frac{1}{(1-(q-1)\beta\mu)^{1/(q-1)-(3/2)}} S_2, \tag{19}$$

such that the temperature dependence of the fugacity $z$ is given by

$$\ln z = \frac{1}{q-1}\left[1-\left(\frac{-2\pi V}{h^3}\left(\frac{2m}{(q-1)\beta}\right)^{3/2} S_2\right)^{2(q-1)/(5-3q)}\right], \tag{20}$$

restricting the fugacity to the interval $0 < z < e^{1/(q-1)}$. An inspection of Eq.(17) shows that the correct way of defining the average energy is through the equation

$$\langle U\rangle = \frac{4\pi V}{h^3}\int_0^\infty \frac{p^2}{2m}\langle n(p)\rangle^q p^2 dp, \tag{21}$$

such that the average energy per particle becomes

$$\frac{\langle U\rangle}{\langle N\rangle} = \frac{kT}{q-1}\frac{\int_0^\infty x^4(1+x^2)^{-q/(q-1)}dx}{\int_0^\infty x^2(1+x^2)^{-1/(q-1)}dx}. \tag{22}$$

Convergence of these integrals restricts the values of $q$ to the interval $1 \le q \le 1.5$ With use of the simple identity

$$\int_0^\infty \frac{d}{dx}\frac{x^3}{(1+x^2)^{1/(q-1)}}dx = 0, \tag{23}$$

it is easily checked that Eq.(21) leads to the required classical result

$$\langle U\rangle = \frac{3}{2}\langle N\rangle kT. \tag{24}$$

Therefore Eq.(21) indicate how we should calculate the average energy for boson and fermion systems such that the classical behavior be recovered in the limit of high temperatures.

In order to calculate some thermodynamic properties of boson and fermion systems with particle occupation numbers given by Eq.(2), we consider the general density operator

$$\hat{\rho} = \frac{1}{Z} \prod_{j=0} \left(1 + \beta(q-1)\hat{K}_j\right)^{1/(1-q)}, \tag{25}$$

with the partition function $Z$ given by the expression

$$Z = \prod_{j=0} \sum_{n_j=0} \left(1 + \beta(q-1)\hat{K}_j\right)^{1/(1-q)}, \tag{26}$$

such that $Tr\hat{\rho} = 1$. We wish to find the form of the function $\hat{K} = \sum_j \hat{K}_j$ which would lead to the average occupation numbers given in Eq (2) and the standard thermodynamics expressions as $q \to 1$. It is natural to assume that the function will depend on the energy $\varepsilon_j$, the chemical potential $\mu$ and the number operator $\hat{n}_j$. Following our procedure for the classical case, we will calculate the occupation number and the average energy from the corresponding definitions $\langle N \rangle = Tr\hat{\rho}\hat{N}$ and $\langle U \rangle = Tr\hat{\rho}^q\hat{H}$.

### B. Boson case

With use of the standard boson algebra

$$a_{k'} a_k^\dagger - a_k^\dagger a_{k'} = \delta_{k,k'}, \tag{27}$$

and assuming that $a_k^\dagger \hat{\rho} = \rho_k \hat{\rho} a_k^\dagger$, it is simple to check that the expectation value of the product of a creation and annihilation operator can be written as

$$\langle a_k^\dagger a_{k'} \rangle = -\delta_{k,k'} + \rho_k \langle a_k^\dagger a_{k'} \rangle, \tag{28}$$

leading to

$$\langle a_k^\dagger a_{k'} \rangle = \frac{\delta_{k,k'}}{\rho_k - 1}, \tag{29}$$

where the function $\rho_k$ is given by Eq.(11). Based on the observation that for an arbitrary function $g$, independent of the operators $a$ and $a^\dagger$, and the number operator $\hat{n}_j = a_j^\dagger a_j$ we have the identity $g^{\hat{n}_j} a_j^\dagger = g a^\dagger g^{\hat{n}_j}$, we find that the function $K_j$ is given by the expression

$$\hat{K}_j = \frac{1}{\beta(1-q)} \left(1 - \rho_j^{(q-1)\hat{n}_j}\right), \tag{30}$$

where we need to restrict $q$ to those values such that $1/(q-1)$ is an integer. Therefore, the parameter $q = (j+1)/j$, where $j$ is a positive integer. For those values of $q$ such that $1/(q-1)$ is an odd number the function $n(\varepsilon) > 0$ for all values of the energy and $\mu \leq 0$. For $1/(q-1)$ even, although $n(0) > 0$ for $\beta\mu \notin [0, 2/(q-1)]$, the derivative $\partial\langle n \rangle/\partial\varepsilon < 0$ for $\mu < 0$ only. Therefore, as in the standard Bose-Einstein case the chemical potential is negative for all temperatures $T > 0$.

The density operator becomes

$$\hat{\rho} = \frac{1}{Z_{BE}} \prod_{j=0} \rho_j^{-\hat{n}_j}, \tag{31}$$

with the partition function $Z$ given by

$$Z_{BE} = \prod_{j=0} \frac{1}{1 - \rho_j^{-1}}. \tag{32}$$

Equation (30) can be rewritten in a more suggestive form by defining the variable $Q_k = \rho_k^{q-1}$ such that the full operator $K$ reads

$$\hat{K} = \sum_{j=0} (\varepsilon_j - \mu) \frac{1 - Q_j^{\hat{n}_j}}{1 - Q_j}, \tag{33}$$

which clearly contains a sum of products of the factor $(\varepsilon_j - \mu)$ times a $Q_j$-number. As a matter of fact, we can rewrite Eq. (33) in a more standard form as follows

$$\hat{K} = \sum_j (\varepsilon_j - \mu) \overline{\phi}_j \phi_j, \tag{34}$$

where the set of operators $\phi_j$ and its adjoint $\overline{\phi}_j$ are defined in terms of boson operators according to

$$\overline{\phi}_j = a_j^{\dagger} \tag{35}$$

$$\phi_j = a_j^{\dagger -1} \frac{1 - Q_j^{\hat{n}_j}}{1 - Q_j}. \tag{36}$$

The operators $\phi_j$ and $\overline{\phi}_j$ satisfy an algebra reminiscent of the so-called $q$-boson algebra, which is

$$\phi_k \overline{\phi}_j - Q_k^{\delta_{j,k}} \overline{\phi}_j \phi_k = \delta_{j,k}. \tag{37}$$

For the average internal energy $\langle U \rangle = Tr \hat{\rho}^q \hat{H}_B$, with $\hat{H}_B = \sum_k \varepsilon_k \overline{\phi}_k \phi_k$, we simply get

$$\langle U \rangle = \sum_k \frac{\varepsilon_k}{\rho_k^q - \rho_k^{q-1}}. \tag{38}$$

In order to evaluate Eqs. (2) and (38) in a simple way we approximate these equations by considering the thermodynamic limit. Therefore we write

$$\langle N \rangle = \langle N(0) \rangle + \frac{V}{\lambda_T^3} G_{3/2}(z, q), \tag{39}$$

$$\langle U \rangle = \frac{3V}{2\beta \lambda_T^3} G_{5/2}(z, q), \tag{40}$$

where we have separated, as usual, the ground state occupation number, $\lambda_T^2 = h^2/2\pi mkT$, and the functions $G_\nu$ are written

$$G_{3/2}(z, q) = \frac{4}{\sqrt{\pi}(q-1)^{3/2}} \int_0^\infty \frac{x^2}{[1 + \Omega]^{1/(q-1)} - 1} dx, \tag{41}$$

$$G_{5/2}(z, q) = \frac{8}{3\sqrt{\pi}(q-1)^{5/2}} \int_0^\infty \frac{x^4}{[1 + \Omega]^{q/(q-1)} - (1 + \Omega)} dx, \tag{42}$$

where $\Omega = x^2 - (q-1)\ln z$. The functions $G_{3/2}(z,q)$ and $G_{5/2}(z,q)$ become for $q=1$ the well known Bose-Einstein functions $g_\nu = \sum_{n=1} z^n/n^\nu$ for $\nu = 3/2, 5/2$ respectively. Figure 1 shows the graphs of the functions $G_{3/2}(1,q)$ and $G_{5/2}(1,q)$ in comparison to the functions $g_{2/3}(1,q)$ and $g_{5/2}(1,q)$ for the quantum group case in Ref. [21]. As expected these functions converge at $q=1$ to the values $\zeta(3/2) = 2.612$ and $\zeta(5/2) = 1.341$ respectively. These graphs indicate that the thermodynamic behavior for the fractal case will be different than the case of quantum group bosons. In particular, since the critical temperature $T_c$ is proportional to $1/G_{3/2}(1,q)^{2/3}$, we see from Figure 1 that the critical temperatures for a Bose gas $T_c^{BE}$ and for a quantum group boson gas $T_c^{QG}$ compare to $T_c$ as follows

$$T_c < T_c^{BE} < T_c^{QG}. \tag{43}$$

Defining the function $\langle u\rangle = (2\pi m/h^2)(V^2/\langle N\rangle^5)^{1/3}\langle U\rangle$ in Eq.(40) we write

$$\langle u\rangle = \frac{3}{2}\frac{G_{5/2}(z,q)}{G_{3/2}^{5/3}(1,q)}\left(\frac{T}{T_c}\right)^{5/2}. \tag{44}$$

Fig. 2 shows the function $\langle u\rangle(T)$ for $q=1.1$ and $q=1.2$, for an ideal gas and quantum group $SU_{1.1}(2)$. From the figure we see that $U^{QG} > U^{BE} > U$ and solely the quantum group case exhibits a discontinuous derivative at the critical temperature. The heat capacity $C_V = (\partial\langle U\rangle\partial T)_V$ is given by

$$C_V(T) = k\langle N\rangle\left[\frac{15}{4}\frac{G_{5/2}(z,q)}{G_{3/2}(z,q)} - \frac{9}{4}\frac{G_{3/2}(z,q)}{G_{3/2}(1,q)}\frac{\frac{\partial G_{5/2}(z,q)}{\partial z}}{\frac{\partial G_{3/2}(z,q)}{\partial z}}\right]\left(\frac{T}{T_c}\right)^{3/2}, \tag{45}$$

which reduces for $T < T_c$, $\mu = 0$ and $\partial\mu/\partial T = 0$ to

$$C_V(T<T_c) = \frac{15}{4}\langle N\rangle k\left(\frac{T}{T_c}\right)^{3/2}\frac{G_{5/2}(1,q)}{G_{3/2}(1,q)}. \tag{46}$$

Since the function $\partial G_{3/2}/\partial z$ is not finite in the $z\to 1$ limit, there is no gap in the heat capacity at $T = T_c$. Figure 3 shows a graph which compares the heat capacity for $q = 1.1$, as given by Eq.(45), with the textbook case of an ideal Bose-Einstein gas and the quantum group case as reported in Ref. [21]. At high temperatures, the heat capacity $C_V$ in Eq.(45) tends to the classical result $C_V = (3/2)k\langle N\rangle$. The entropy function $S = ((1/\beta)\ln Z + \langle U\rangle) - \mu\langle N\rangle)(1/T)$ is given by the expression

$$S = \frac{k\langle N\rangle}{G_{3/2}(1,q)}\left[\frac{3G_{5/2}(z,q)}{2} - \frac{4}{(q-1)^{3/2}\sqrt{\pi}}\int_0^\infty x^2\ln\left(1-(1+\Omega)^{1/(1-q)}\right)dx - \ln z\right]\left(\frac{T}{T_c}\right)^{3/2}. \tag{47}$$

Figure 4 shows a graph of the entropy per particle for $q = 1.1$ in comparison to those corresponding to the ideal Bose gas and the $SU_{1.1}(2)$ cases. For the quantum group $SU_{1.1}(2)$ the entropy is given by the expression

$$S^{QG} = \frac{5}{2}k\langle N\rangle\frac{g_{5/2}(z,q)}{g_{3/2}(1,q)}\left(\frac{T}{T_c^{QG}}\right)^{3/2} - \ln z. \tag{48}$$

From these graphs we see that the behavior near the corresponding critical temperature is very different than the quantum group case in which the heat capacity exhibits a gap that increases with the value of the parameter $q$ and the entropy has a discontinuous derivative.

### C. Fermion case

For the fermion case it is simple to check that

$$\prod_j \left(1+\beta(q-1)\hat{F}_j\right) b_k^\dagger = (1+\beta(q-1)(\varepsilon_k-\mu))b_k^\dagger \prod_j \left(1+\beta(q-1)\hat{F}_j\right), \tag{49}$$

where the operator $\hat{F}_j = (\varepsilon_j-\mu)b_j^\dagger b_j$. Therefore the occupation number $\langle b_j^\dagger b_j\rangle = Tr\hat{\rho} b_j^\dagger b_j$ and the average internal energy $\langle U\rangle = Tr\hat{\rho}^q \hat{H}_F$ become

$$\langle N\rangle = \sum_j \frac{1}{1+(1+\beta(q-1)(\varepsilon_j-\mu))^{1/(q-1)}} \tag{50}$$

$$\langle U\rangle = \sum_j \frac{\varepsilon_j}{1+(1+\beta(q-1)(\varepsilon_j-\mu))^{q/(q-1)}}. \tag{51}$$

We consider, as in the previous section, the thermodynamic limit of Eqs.(50) and (51)

$$\langle N\rangle = \frac{4V}{\sqrt{\pi}}\left(\frac{1}{\sqrt{q-1}\lambda_T}\right)^3 F_{3/2}(z,q), \tag{52}$$

$$\langle U\rangle = \frac{4V}{\sqrt{\pi}}\frac{1}{\beta}\frac{1}{(q-1)^{5/2}}\frac{1}{\lambda_T^3}F_{5/2}(z,q) \tag{53}$$

where the functions $F_\nu(z,q)$ are defined by

$$F_{3/2} = \int_0^\infty \frac{x^2}{1+[1+\Omega]^{1/(q-1)}}dx, \tag{54}$$

$$F_{5/2} = \int_0^\infty \frac{x^4}{1+[1+\Omega]^{q/(q-1)}}dx. \tag{55}$$

Performing a similar calculation already done in the previous section, we find for the heat capacity $C_V$

$$C_V = k\langle N\rangle \frac{1}{(q-1)^{5/2}}\left(\frac{15}{2}F_{5/2}(z,q) - \frac{9}{2}F_{3/2}(z,q)\frac{\frac{\partial F_{5/2}(z,q)}{\partial z}}{\frac{\partial F_{3/2}(z,q)}{\partial z}}\right)\left(\frac{kT}{\mu_0^{(f)}}\right)^{3/2}. \tag{56}$$

where $\mu_0^{(f)} = (3\langle N\rangle/4\pi V)(h^2/2m)^{3/2}$ is the Fermi energy. For the quantum group $SU_q(2)$, the heat capacity is given by the equation

$$C_V^{QG} = k\langle N\rangle\left[\frac{15}{2}f_{5/2}(z,q) - \frac{9}{2}f_{3/2}(z,q)\frac{\frac{\partial f_{5/2}(z,q)}{\partial z}}{\frac{\partial f_{3/2}(z,q)}{\partial z}}\right]\left(\frac{kT}{\mu_0^{(f)}}\right)^{3/2}, \tag{57}$$

where the functions $f_{3/2}$ and $f_{5/2}$ are given as follows

$$f_{3/2}(z,q) = \int_0^\infty \frac{x^2\left(1+ze^{-q^{-2}x^2}\right)}{f}dx, \tag{58}$$

$$f_{5/2}(z,q) = \int_0^\infty \frac{x^4\left(2+(q^{-2}+1)e^{-q^{-2}x^2}\right)}{f}dx, \tag{59}$$

with $f = e^{x^2}z^{-1}+2+e^{-q^{-2}x^2}z$. Figure 5 shows a graph of the heat capacity for $q=1.1$ and $q=1.2$, the ideal Fermi gas and the quantum group $SU_{1.1}(2)$ fermion case as a function of the temperature. In general, the graphs show that the heat capacities follows the relation

$$C_V < C_V^{Fermi} < C_V^{QG} \tag{60}$$

## III. CONCLUSIONS

In this paper we have studied the thermodynamics of the fractal distribution functions for Maxwell-Boltzmann, Bose-Einstein and Fermi systems in the thermodynamic limit. As pointed out in the Introduction, the parameter $q$ measures the departure from the Boltzmann-Gibbs statistics. As our starting point, we require that the formalism be based on a factorized probability density and be consistent with the classical limit. Based on this, with use of the operator formalism we obtained the matrix density operators that lead to the quantum generalized particle distributions $n(\varepsilon)$. For the Bose-Einstein case, we obtained the interesting result that the hamiltonian can be interpreted as a $q$-deformation of the boson ideal gas hamiltonian. We should remark that for the boson case our partition function differs from the factorized partition function considered in Ref. [9]. A calculation of the heat capacity that results from the boson generalized distribution function shows that it is continuous at the critical temperature. This contrasts with the quantum group case wherein the heat capacity exhibits a discontinuity at that temperature that increases with the value of the deformation parameter. For the fermion case, the heat capacity function for the Fermi ideal gas lies between the heat capacity for the fractal and quantum group cases. In general, our results point out that the thermodynamic behavior given by both fractal distribution functions are very different to those found for Bose-Einstein and Fermi systems with quantum group symmetry. Since, as claimed for several authors, the fractal Bose-Einstein and Fermi distribution functions approximate well, under some conditions, the nonextensive behavior of some physical systems, the relation that the latter may have with quantum group symmetric systems seems to be rather limited to the context of $q$-analysis.

---

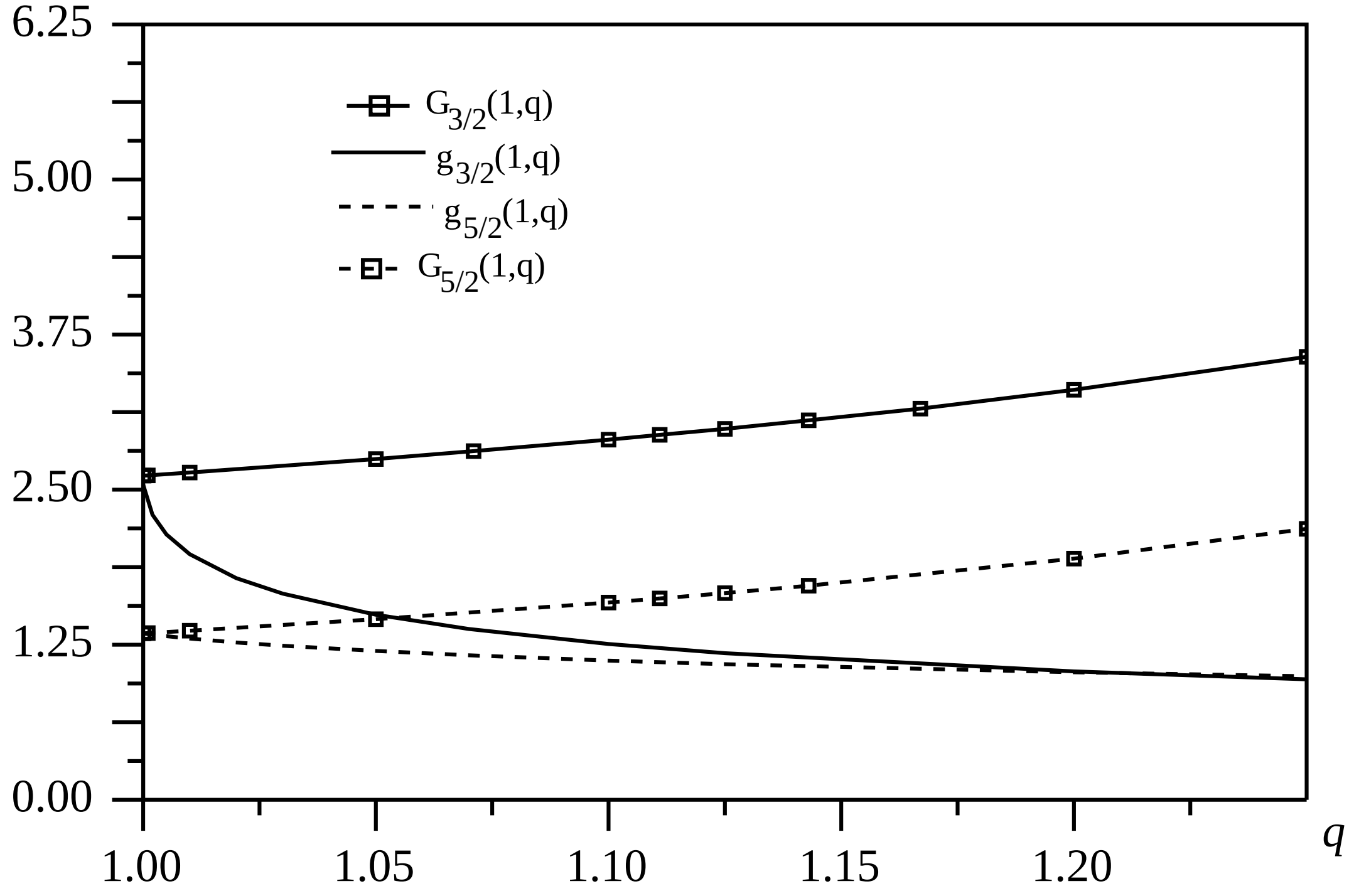

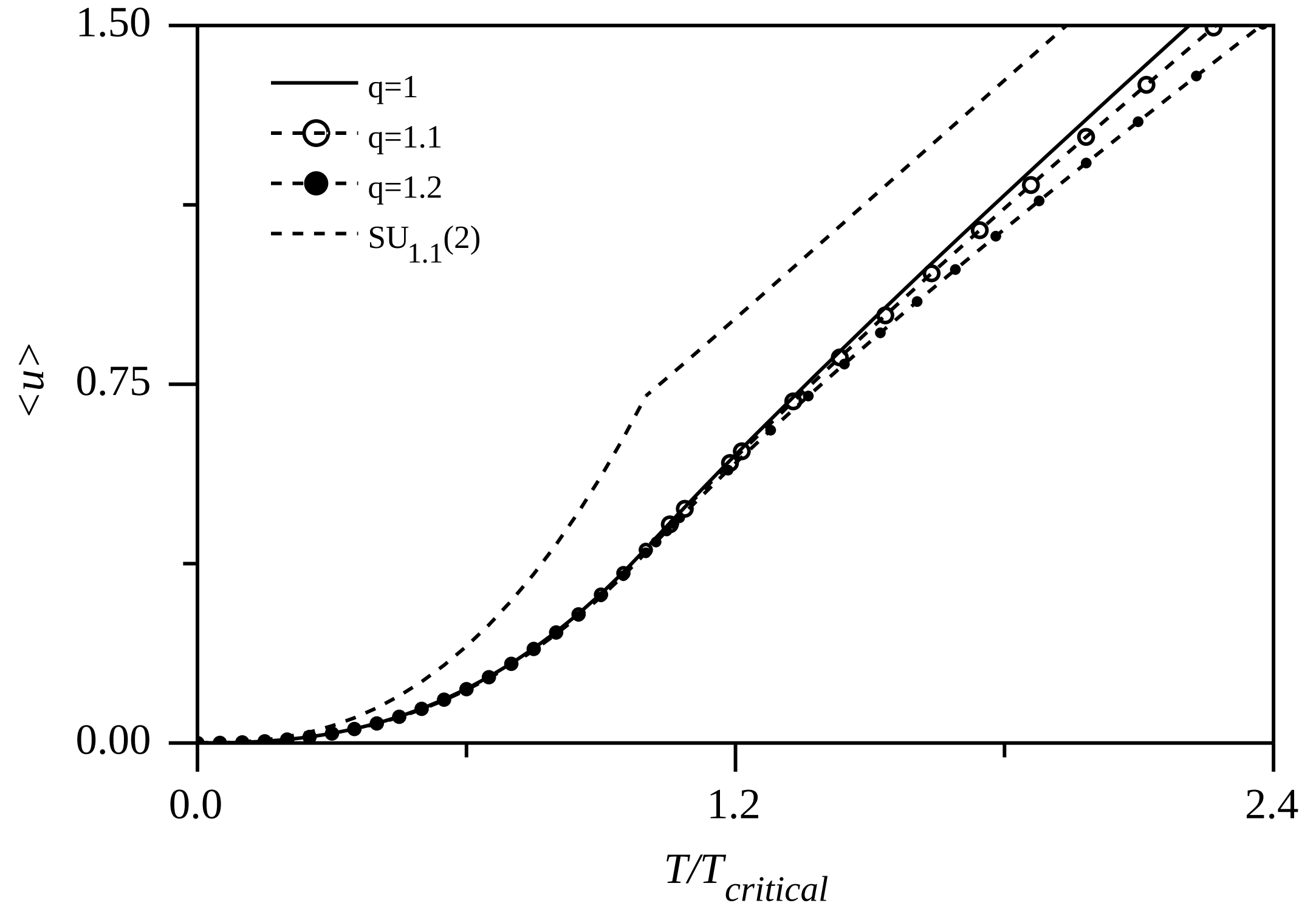

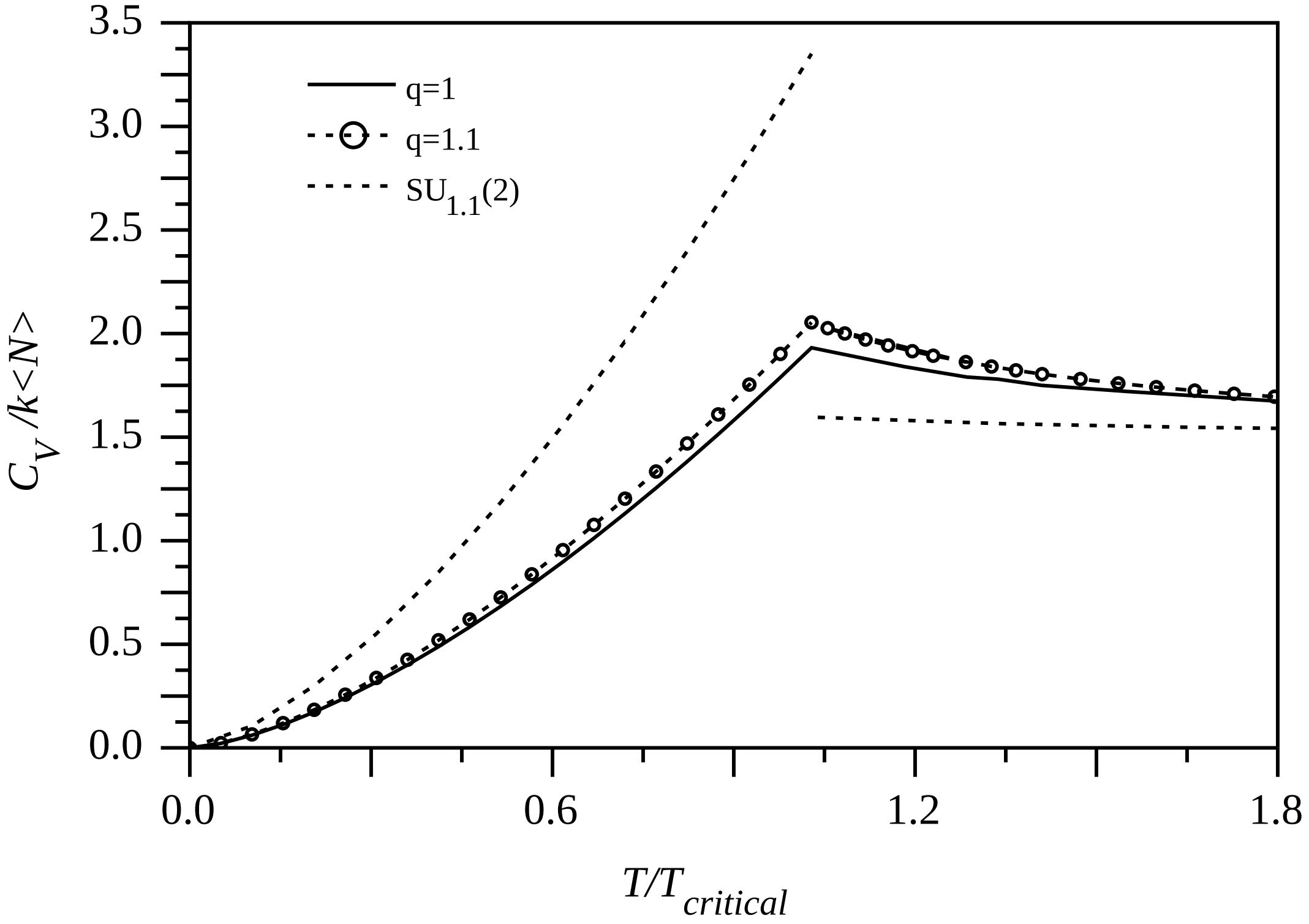

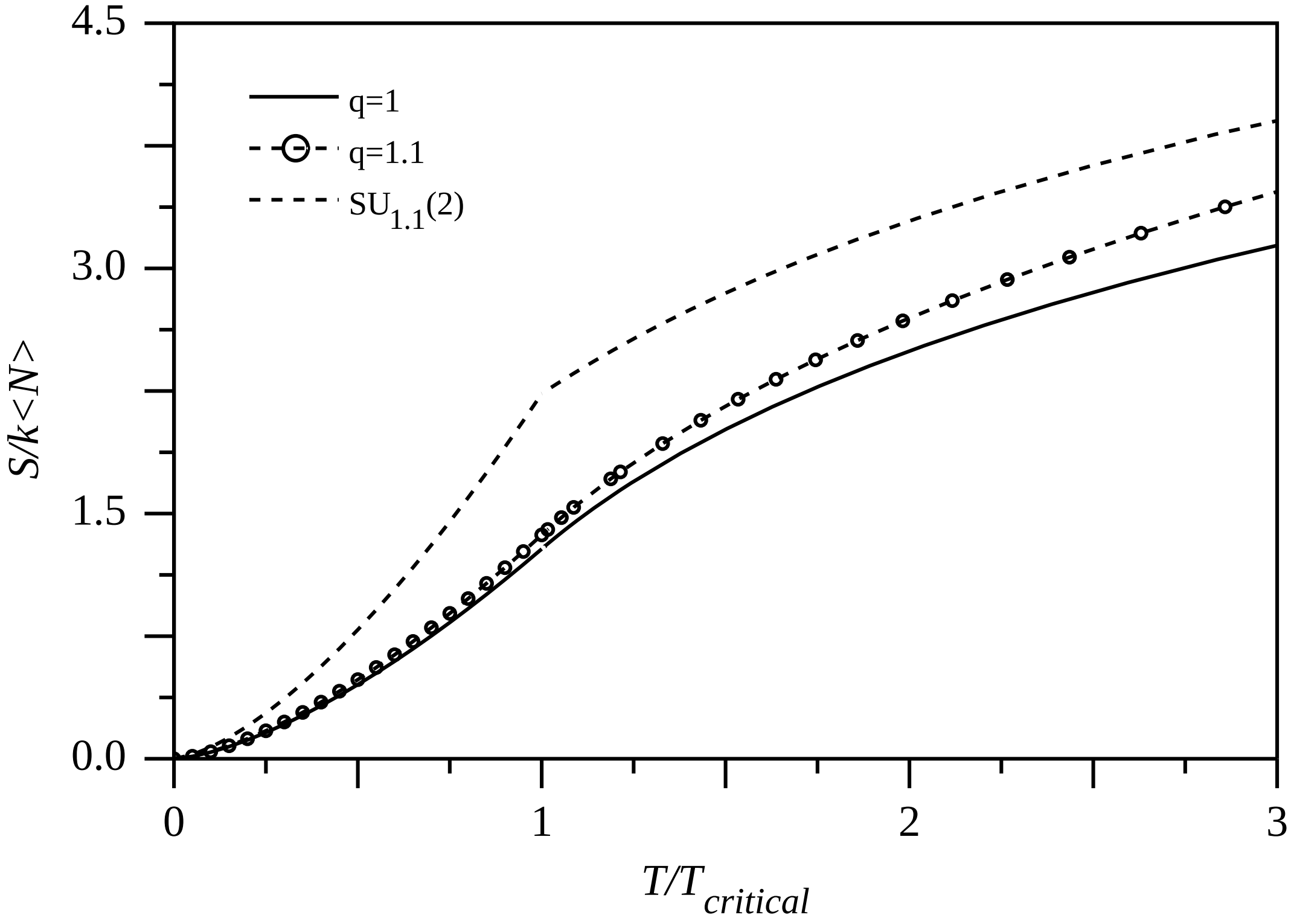

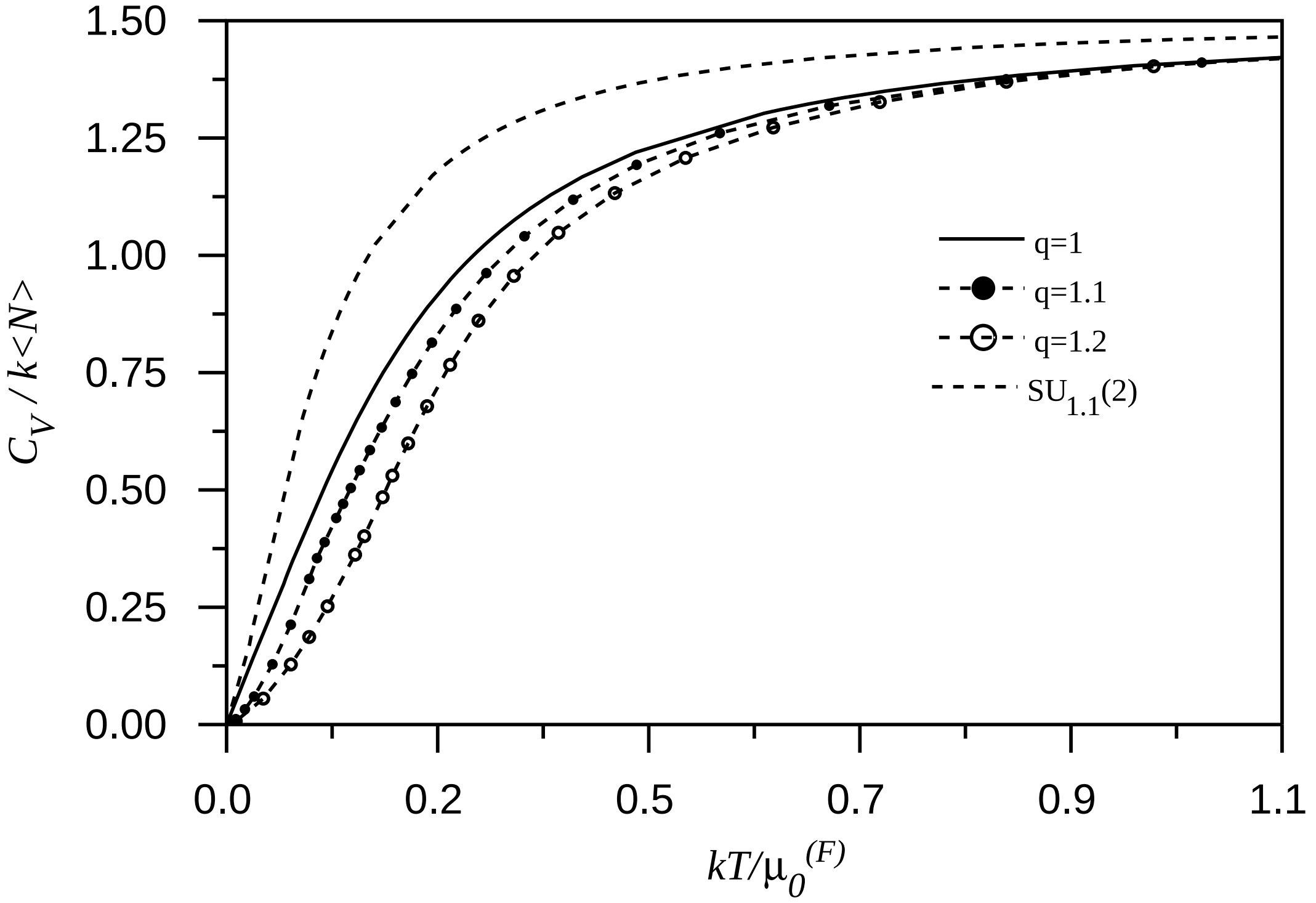

FIG. 1. The functions $G_{3/2}(1,q)$ and $G_{5/2}(1,q)$ as defined in the text, and the functions $g_{3/2}(1,q)$ and $g_{5/2}(1,q)$ from Ref. [21] for the quantum group, $SU_q(2)$, case for the interval $1 < q < 1.25$.

FIG. 2. The function $u$, as defined in the text, for $q = 1.1$ and $q = 1.2$, for an ideal Bose-Einstein gas, and for the $SU_{1.1}(2)$ boson case as functions of $T/T_{critical}$, where $T_{critical}$ refers to the critical temperatures $T_c$, $T_c^{BE}$ and $T_c^{QG}$, respectively.

FIG. 3. The heat capacity $C_V$ as a function of $T/T_{critical}$ showing the sharp difference between the phase transition for the quantum group case in comparison to the Bose-Einstein and fractal cases.

FIG. 4. The entropy function $S$ for the fractal $q = 1.1$, Bose-Einstein and $SU_{1.1}(2)$ cases.

FIG. 5. The heat capacity as a function of the temperature for $q = 1.1$ and $q = 1.2$, the ideal Fermi gas and the quantum group $SU_{1.1}(2)$ fermion cases.